\documentclass[aps,prb,twocolumn,superscriptaddress,showpacs]{revtex4}
\usepackage{graphicx}

\begin{document}


\title{Absence of magnetic phase separation in MnSi under pressure}


\author{D.~Andreica}
\affiliation{Laboratory for Muon-Spin Spectroscopy, Paul Scherrer Institute, 5232 Villigen-PSI, 
Switzerland}
\affiliation{Faculty of Physics, Babes-Bolyai University, 400084 Cluj-Napoca, Romania}
\author{P.~Dalmas de R\'eotier}
\affiliation{CEA/DSM/Institut Nanosciences et Cryog\'enie, 38054 Grenoble, France}
\author {A.~Yaouanc}
\affiliation{CEA/DSM/Institut Nanosciences et Cryog\'enie, 38054 Grenoble, France}
\author{A.~Amato}
\affiliation{Laboratory for Muon-Spin Spectroscopy, Paul Scherrer Institute, 5232 Villigen-PSI, 
Switzerland}
\author {G.~Lapertot}
\affiliation{CEA/DSM/Institut Nanosciences et Cryog\'enie, 38054 Grenoble, France}

\author{}
\affiliation{}



\date{\today}

\begin{abstract}

We report muon spin spectroscopy data ($\mu$SR) obtained under hydrostatic 
pressure on a large single crystal of the itinerant 
helimagnet MnSi, and recorded down to $0.235 \, {\rm K}$ and up to 
$15.1 \, {\rm kbar}$. Up to the critical pressure
$p_c = 14.9 \, (2)$~kbar, where the magnetic order is suppressed, the 
$\mu$SR data unambiguously demonstrate that the ground state of the system is 
magnetic with no indication of any phase separation.   

\end{abstract}


\maketitle

The strongly correlated intermetallic cubic compound MnSi has been known for a long time to 
order magnetically below $T_c  \simeq 29 \, {\rm K}$ in a long period helical 
structure. \cite{Williams66,Ishikawa76} 
The first-order nature of the transition has only been demonstrated quite 
recently.\cite{Stishov08,Petrova09a} The study of the effect of pressure 
on the magnetic phase has 
shown that the transition temperature decreases with pressure, remaining first order, and tending to zero at 
$p_c \simeq 14 \, {\rm kbar}$. \cite{Thompson89,Petrova09b}

It is well known that phase separation might occur near the temperature of a first order-phase transition.
Previous muon spin rotation and relaxation  ($\mu$SR) experiments by Uemura and coworkers indicated a Magnetic Phase Separation (MPS)
in MnSi.\cite{Uemura07} Prior to the $\mu$SR work, a Nuclear Magnetic Resonance (NMR) study suggested 
MPS in the same pressure range $\Delta p \simeq 3 \, {\rm kbar}$ just below $p_c$.\cite{Yu04} However, since 
the NMR sample had to be powderized, strains problems could induce extrinsic 
MPS. Hence, probing a single crystal, as it has been done in Ref.~\onlinecite{Uemura07}, is certainly an
advantage. However, the validity of the conclusions drawn from these measurements 
can be questioned since, according to Fig. 2c of Ref.~\onlinecite{Uemura07}, solely one spontaneous 
frequency was detected below $T_c$ rather than two, as expected from previous ambient pressure 
measurements.\cite{Kadono90} In addition, the measurements were performed only down to 
$2.5 \, {\rm K}$, which might be an issue when studying the behaviour of the system
near $p_c$ where $T_c$ goes to 0.

To clarify such issues, we have carried out a series of accurate $\mu$SR experiments under pressure down 
to $0.235 \, {\rm K}$ on a large MnSi single 
crystal. The main result of our study is that, up to $p_c$ and at low temperatures, the full sample volume becomes
magnetic. Therefore, despite the first order nature of the phase transition 
that is clearly confirmed by our data, no phase separation is detected in MnSi.

Due to its unique sensitivity to slow spin fluctuations and its local probe character providing the 
possibility to distinguish paramagnetic 
and magnetically ordered volume fractions,\cite{Dalmas97,Amato97,Dalmas04,Amato04} no matter the compound chemical 
composition, the $\mu$SR spectroscopy is well suited to probe whether MPS is effectively an intrinsic property of 
MnSi. 
Since our main interest was on the existence of MPS, most of the $\mu$SR measurements were performed with 
the so-called Weak-Transverse-Field (WTF) method, which is well adapted to determine the magnetic 
volume fraction. A small field, $B_{\rm ext} = 5 \, {\rm mT}$ in our case, is applied perpendicular 
to the muon polarization ${\bf S}_\mu$. 
The depolarization function for the muons stopping in the sample is then the sum of two precessing components. While the one which arises
from the paramagnetic volume precesses at a frequency associated to the 
applied field and is weakly damped, the other one related to
muons stopped in the magnetically ordered phase, precesses at a frequency
corresponding to the spontaneous field and exhibits a strong depolarization. 
This latter damping stems from the relatively large 
magnetic field distribution at the muon site inherent to an ordered magnetic 
state. By carefully determining the amplitude of these components, the volume 
ratio between the magnetic and paramagnetic
phases can be accurately determined as a function of temperature and pressure. 
Measurements in zero applied magnetic field were also performed at different 
pressures.   
 
The $\mu$SR measurements were carried out at the 
General Purpose Decay-Channel (GPD) spectrometer of the Swiss Muon Source 
(S$\mu$S, Paul Scherrer Institute, Villigen, Switzerland).

A detailed description of the pressure cell used for the $\mu$SR measurements is given
elsewhere.\cite{Andreica01,Serdar10} Relative to previous works, 
some improvements were made. The pressure cell was mounted either on the sample stick of a Janis $^4$He-flow cryostat
or on the cold finger of an 
Oxford Instrument $^3$He cryostat. The temperature range from $40 \, {\rm K}$ down to $0.235 \,{\rm K}$ could hence
be covered. The extension to low 
temperature by 
more than an order of magnitude relative to previous $\mu$SR work \cite{Uemura07} is a key ingredient to unravel the
new features reported in the present study. For each pressure in this study, the exact pressure applied on
the sample was measured {\sl in situ} at low temperature using the pressure dependence of the superconducting 
transition temperatures of small pieces of either In or Pb placed inside the pressure cell. 
It is worth mentioning that the pressure determination was performed at similar temperatures as 
the $\mu$SR measurements themselves. The pressure transmitting medium was a 1:1 mixture of n-pentane and isoamyl alcohol which is known for the excellent hydrostaticity conditions it provides in our pressure range of interest and for the absence of hysteresis effects.\cite{Butch09}
 
The MnSi sample was a cylinder of $7 \, {\rm mm}$ diameter and $19 \, {\rm mm}$ length cut from a single 
crystal prepared similarly as the one used in previously published ambient pressure $\mu$SR measurements,\cite{Yaouanc05}
elastic neutron scattering investigation under pressure\cite{Fak05} and thermal expansion studies\cite{Miyake09}
under pressure. The single crystal was grown by the Czochralsky pulling technique from a stoichiometric 
melt of high purity elements ($>99.995 \, \%$) using radio-frequency heating and a cold copper crucible. 
The residual resistivity ratio of such prepared crystals is about 40. Scanning electron microscope 
microanalysis and backscattered electron images reveal neither any deviation 
from the known crystal structure nor any presence of foreign phase.
The possibility for off-stoichiometry was carefully investigated by examining 
the polycrystalline ingot remaining after a pulling especially 
designed to consume 99~\% of the initial load, instead of $\simeq$ 25~\% for 
normal growth. No sizeable fraction of any foreign phase was found in the 
remaining
ingot, confirming a stoichiometry in the ratio 1:1 in our MnSi single crystal.

\begin{figure}
\center 
\includegraphics[scale=0.75]{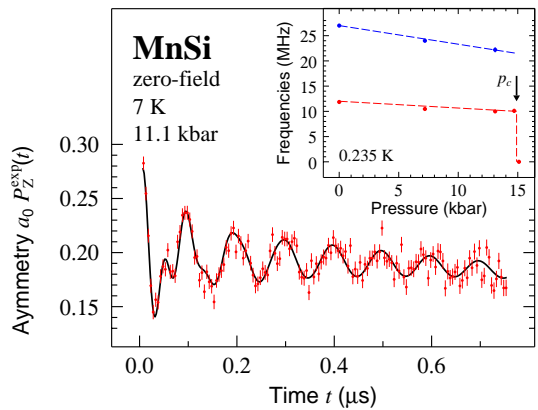} 
\caption{\label{MnSi_pressure_spectrum_2freq}(color online). 
An example of a zero-field $\mu$SR spectrum recorded on a single crystal of MnSi at 7~K. 
The pressure as determined by the superconducting temperature of indium is $11.1 \, {\rm kbar}$. 
The solid line is the result of the fit consisting of two spontaneously
precessing components plus an exponential spin-lattice relaxation
one. The oscillations are
 described by exponentially damped cosine functions with frequencies 
$\nu_1 = 9.99 \, (3)  \, {\rm MHz}$ and 
$\nu_2 = 22.9 \, (5)  \, {\rm MHz}$ and an amplitude ratio similar to the one 
reported at ambient pressure value.\cite{Kadono90} The amplitude of the 
spin-lattice relaxation 
channel is found to account for one third of the total MnSi signal, a ratio 
expected for a weakly anisotropic cubic magnet like MnSi irrespective of the
orientation of the muon initial polarization relative to the crystal axes.
The relaxation rate is $\lambda_Z = 0.11 \, (5) \, 
\mu{\rm s}^{-1}$. The signal 
from the pressure cell is characterized and described in Refs.~\onlinecite{Andreica01} and
\onlinecite{Serdar10}.
Nearly half of the measured signal originates from the sample, a much larger
proportion than the $\sim 30 \, \%$ of the published work.\cite{Uemura07}
The insert displays the pressure dependence of the two spontaneous precession
frequencies measured at 0.235~K. At 14.7~kbar, the upper one is too heavily 
damped to be realiabily extracted from the spectrum. The dashed lines are 
guides to the eye.}
\end{figure}

In Fig.~\ref{MnSi_pressure_spectrum_2freq} a zero-field $\mu$SR spectrum 
recorded at $7 \, {\rm K}$ under $11.1 \, {\rm kbar}$ is reported. It serves 
to display the sample 
and pressure quality. The beating of the expected two oscillating components
is clearly seen. Had the field distribution in the sample been large, due for 
example to a large pressure gradient or sample inhomogeneity, the higher 
frequency would not have been detected. 
Note that this higher frequency precession was not observed in the 
previous $\mu$SR work under pressure.\cite{Uemura07}
The pressure dependence of the two low temperature 
spontaneously precessing frequencies,
which reflect the pressure dependence of the order parameter, is
displayed in the insert of Fig.~\ref{MnSi_pressure_spectrum_2freq}. 
An abrupt change is observed between 14.7 and 15.1~kbar, confirming
the first order nature of the phase transition. 
Belitz and coworkers predicted this type of transition to 
occur in weak ferromagnets.\cite{Belitz99}

\begin{figure} 
\center 
\includegraphics[scale=0.75]{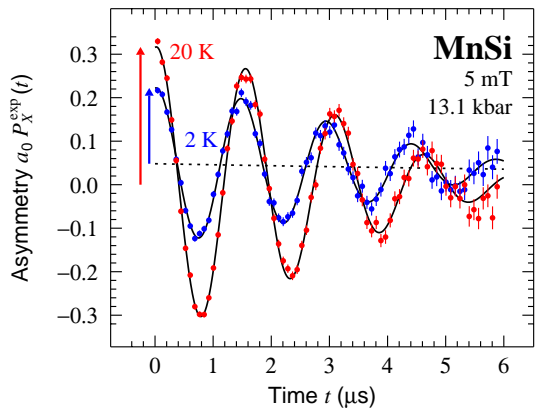}
\caption{\label{MnSi_pressure_wtf_spectra}(color online). 
Spectra recorded in a weak transverse field of 5~mT in MnSi under a pressure
of 13.1~kbar. Note that the time scale differs completely with
the one of Fig.~\ref{MnSi_pressure_spectrum_2freq}. The binning of the data
which is adopted here averages out the oscillations of the muon spin
in the MnSi spontaneous field. The initial amplitude of the 20~K 
spectrum which is marked by the longer arrow 
has the expected value in our experimental conditions. At 2~K, the precessing
component observed in the plot corresponds to muons which are stopped in the 
pressure cell: its amplitude is indicated by the shorter arrow. The 
spin-lattice relaxation signature of the muons which are implanted in MnSi is 
seen as a shift in the mean value of the precessing signal.
The solid lines are fits to a product of a cosine function
and a Gaussian damping, with an additional 
exponential spin-lattice relaxation function (shown as a dotted line)
for the 2~K spectrum.
The Gaussian damping corresponds to a dominating inhomogeneous broadening.}
\end{figure}

\begin{figure} 
\center 
\includegraphics[scale=0.10]{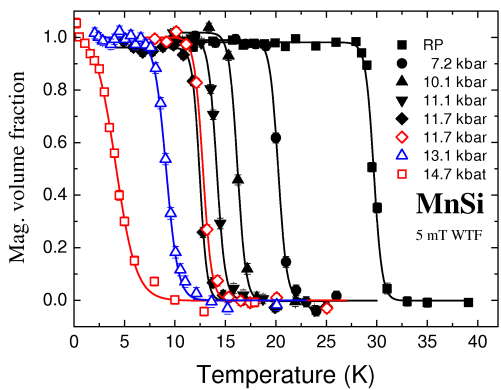}
\caption{\label{MnSi_pressure_vol_vs_temp}(color online). 
Temperature dependence of the magnetic volume fraction for a crystal of MnSi 
at different applied pressures. 
The data were obtained from $5 \, {\rm mT}$ transverse-field measurements. 
The solid  lines are fits to the phenomenological law given in the main text.
RP stands for room pressure.}
\end{figure}

\begin{figure} 
\center 
\includegraphics[scale=0.75]{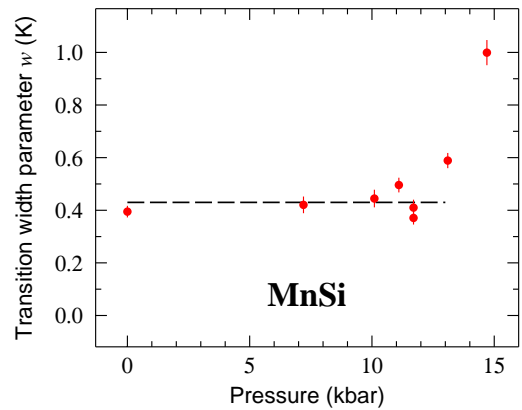}
\caption{\label{MnSi_pressure_transition_width}(color online). 
Pressure dependence of the $w$ parameter determined from the fit of the data in
Fig.~\ref{MnSi_pressure_vol_vs_temp}. As shown by the dashed line, $w$ is 
merely unchanged up to about 13~kbar.
}
\end{figure}

Examples of WTF spectra are shown in Fig.~\ref{MnSi_pressure_wtf_spectra}.
Muons stopped both in the pressure cell and in the MnSi crystal contribute
to them. The muons implanted in paramagnetic MnSi and in the (non magnetic) 
pressure cell precess at a frequency which differs from 
$\gamma_\mu B_{\rm ext}/(2\pi)$ only by a small Knight shift; $\gamma_\mu$
= 851.6 Mrad.s$^{-1}$.T$^{-1}$ is the muon gyromagnetic ratio.
Muons stopped in magnetically ordered MnSi precess at a much higher frequency
(see Fig.~\ref{MnSi_pressure_spectrum_2freq} and note the time scale).
Since at low temperature and room pressure it is known that the whole
volume of MnSi is magnetically ordered, it is simple matter to derive the
pressure cell contribution to the signal. It is then possible to determine
the MnSi magnetic volume fraction. The results are shown in 
Fig.~\ref{MnSi_pressure_vol_vs_temp} as a function of temperature and
pressure.  
The most obvious feature is that the whole sample volume is magnetically 
ordered at low temperature for all the applied pressures. The magnetic
volume fraction is fitted to the phenomenological law
$\{1+\exp[(T-T_c)/w]\}^{-1}$. With this formula, for a given pressure, 
$T_c$ is the temperature at which half of the sample volume is magnetic.
The pressure dependence of $T_c$ will be shown and compared to the
literature results in 
Fig.~\ref{MnSi_pressure_vol_criticaltemp_vs_temp}. 
Up to 13~kbar, the width of the transition observed in 
Fig.~\ref{MnSi_pressure_vol_vs_temp} is relatively narrow, 
pressure independent and about two times narrower than previously reported.
\cite{Uemura07} The data taken at $14.7 \, {\rm kbar}$ single out themselves: 
the transition width is more pronounced.  
This trend if confirmed in Fig.~\ref{MnSi_pressure_transition_width} by the 
plot of the fit parameter $w$ which is proportional to the width of the 
magnetic transition.

\begin{figure} 
\center 
\includegraphics[scale=0.30]{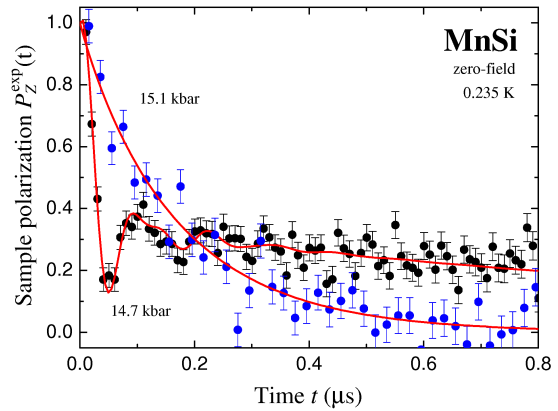}
\caption{\label{MnSi_pressure_spectrum_nearcritical}(color online). 
Time dependence of the zero-field muon polarization for MnSi 
recorded at $0.235 \, {\rm K}$ on each side of the critical pressure, after subtraction of the 
$\mu$SR signal arising from the pressure cell.  
The solid lines represent the best fits to the spectra. The sample depolarization function has an 
exponential character at $15.1 \, {\rm kbar}$. At $14.7 \, {\rm kbar}$ it is the sum of a 
relaxing and of two damped oscillating signals.
}
\end{figure}

\begin{figure} 
\center 
\includegraphics[scale=0.75]{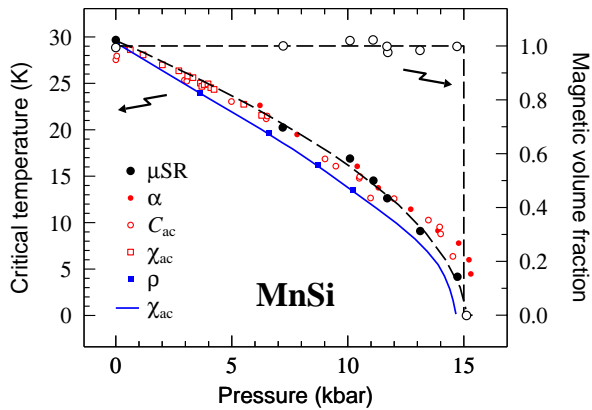}  
\caption{\label{MnSi_pressure_vol_criticaltemp_vs_temp}(color online). 
Low temperature magnetic volume fraction and magnetic critical 
temperature as a function of the applied pressure determined from
our $\mu$SR study.
The dashed lines are guide to the eyes.
The data of the pressure dependence of the critical
temperature are from thermal expansion ($\alpha$),\cite{Miyake09} ac-specific 
heat ($C_{\rm ac}$),\cite{Miyake09} 
ac-susceptibility ($\chi_{\rm ac}$),\cite{Pfleiderer97,Petrova06}
and resistivity ($\rho$).\cite{thessieu95}
}
\end{figure}

We present in Fig.~\ref{MnSi_pressure_spectrum_nearcritical} 
two zero-field spectra recorded at 14.7 and 15.1~kbar, respectively.
The pressure at which the spontaneous $\mu$SR frequencies disappear from the $\mu$SR spectra at low temperatures is 
taken as the critical pressure $p_c$, where the magnetic ground state is suppressed.

The pressure dependence of the low temperature magnetic volume fraction derived from our measurements 
as well as of the critical temperature are plotted in 
Fig.~\ref{MnSi_pressure_vol_criticaltemp_vs_temp}.
The pressure dependence of the critical temperature is consistent with 
recently published results obtained by various techniques (thermal expansion, specific heat 
and ac-susceptibility) up to about $13 \, {\rm kbar}$. The possible discrepancy at higher
pressure (although the experimental points are somewhat distributed) 
may be due to the effect of different pressure media.\cite{Miyake09}

It may be that the sample and pressure quality plays a role in the appearance of MPS. 
A very clear example is given by the research on 
URu$_2$Si$_2$; see Ref.~\onlinecite{Amitsuka07} and references therein. Better 
sample and pressure quality results in a sharper transition in 
URu$_2$Si$_2$, {\em i.e.} no MPS is observed. 

Finally we interpret the data recorded in the vicinity of $p_c$.
 
We first estimate a bound on the characteristic fluctuation time for the spin dynamics under 
$14.7 \, {\rm kbar}$ at $0.235 \, {\rm K}$. Since we clearly observe oscillations up to 
$\sim 0.2 \, \mu{\rm s}$, see  Fig.~\ref{MnSi_pressure_spectrum_nearcritical}, the 
two spontaneous fields do not flip in this time range.\cite{Dalmas06}
Therefore the time characterizing the fluctuations of the two spontaneous fields is at least
as long as $0.2 \, \mu{\rm s}$.

We now discuss the pressure inhomogeneity in our measurements. It seems that a pressure 
distribution of $0.5 \, {\rm kbar}$ is a reasonable value.\cite{Ruetschi07,Miyake09}
Because the depolarization function measured at $0.235 \, {\rm K}$
under $15.1 \, {\rm kbar}$ is fully characteristic of a paramagnetic state and the one 
recorded under $14.7 \, {\rm kbar}$ at the same temperature reflects a magnetic state, we infer 
$p_c = 14.9 \, (2) \, {\rm kbar}$. Since we have now some insight on the quality of the 
applied pressure, we can interpret 
the anomalous large width of the temperature dependence of the magnetic volume fraction 
observed at $14.7 \, {\rm kbar}$; see Fig.~\ref{MnSi_pressure_vol_vs_temp}. 
We expect the $\mu$SR spectra to depend strongly on the pressure near $p_c$. 
In fact, we note that the 
width at $14.7 \, {\rm kbar}$ in our case has about the value found 
by Uemura and collaborators\cite{Uemura07} at low pressure for which they did not observe MPS.

In conclusion, our work shows the importance of pressure 
homogeneity and access to low temperatures for getting reliable information on the phase separation problem.  
We have shown that no MPS is present in MnSi at low temperature. 
Our data are consistent with the existence at low temperature
of a sharp transition at the critical pressure $p_c = 14.9 \, (2) \, {\rm kbar}$.
Pfleiderer and collaborators (see also Ref.~\onlinecite{Fak05}) have recently argued 
for a partial magnetic order above $p_c$.\cite{Reznik04} We expect that $\mu$SR should
be able to study its dynamics. 

We acknowledge useful conversations with B. F{\aa}k.
This research project has been partially supported by the European Commission 
under the 6th Framework Programme through the Key Action: Strengthening the European Research 
Area, Research Infrastructures (Contract number: RII3-CT-2003-505925).
DA acknowledges financial support from the Romanian CNCSIS project 444/2009.
Three of us (PDR, AY and AA) were partially supported by 
the ``Programme d'action integr\'ee PAI franco-suisse Germaine de Sta\"el''. 
Part of this work was performed at the Swiss Muon Source, Paul Scherrer 
Institute, Villigen, Switzerland.

\bibliography{reference.bib}

\end{document}